\documentclass[12pt,preprint]{aastex}
\received{}
\revised{}
\accepted{}

\shorttitle{High $\Cn$ Abundance in Galactic Center Molecular Clouds}
\shortauthors{Tanaka et al.}
\tighten

\newcommand\pcc{\mathrm{ cm^{-3}}}
\newcommand\psc{\mathrm{ cm^{-2}}}
\newcommand\kelvin{\mathrm{ K}}
\newcommand\kmps{\mathrm{ km\,s^{-1}}}
\newcommand\erg{\mathrm{ erg}}
\newcommand\str{\mathrm{ str}}
\newcommand\second{\mathrm{ s}}

\newcommand\Msol{{M_\odot}}

\newcommand\CI{\ion{C}{1}}

\newcommand\Ct{^{13}\mathrm{C}}

\newcommand\Tkin{{T_{\rm kin}}}

\newcommand\nH{{n_{\rm H}}}

\newcommand\Column[1]{{N_{#1}}}

\newcommand\vlsr{{v_{\rm LSR}}}
\newcommand\Tmb{{T_{\mathrm{MB}}}}

\newcommand\TCI{T_{\mathrm{[CI]}}}
\newcommand\TCOt{T_{\mathrm{^{13}CO}}}

\newcommand\CO{{\mathrm{CO}}}

\newcommand\HCNt{{\mathrm{H{^{13}C}N}}}

\newcommand\Cn{{\mathrm{C^0}}}

\newcommand\COt{{\mathrm{{^{13}CO}}}}

\newcommand\NNHp{{\mathrm{N_2H^+}}}

\newcommand\JJ[2]{\mbox{{\it J}=#1\mbox{--}#2}}

\newcommand\CIa{{^3}P_1\mbox{--}{^3}P_0}

\newcommand\gl{l}
\newcommand\gb{b}
\newcommand\NCn{\Column{\Cn}}
\newcommand\NCO{\Column{\mathrm{CO}}}

\newcommand\CIIa{{^2}P_{3/2}\mbox{--}{^2}P_{1/2}}

\usepackage{float}
\begin{document}


\title{High Atomic Carbon Abundance in Molecular Clouds in the Galactic Center Region}
\author{Kunihiko Tanaka}
\email{ktanaka@phys.keio.ac.jp}
\author{Tomoharu Oka}
\author{Shinji Matsumura}
\affil{Department of Physics, Faculty of Science and Technology, Keio University, 3-14-1 Hiyoshi, Yokohama, Kanagawa 223-8522 Japan}

\author{Makoto Nagai}
\affil{High Energy Accelerator Research Organization, 1-1 Oho, Tsukuba, Ibaraki 305-0801 Japan}

\and 

\author{Kazuhisa Kamegai}
\affil{Institute of Space and Astronautical Science, Japan Aerospace Exploration Agency, 3-1-1 Yoshinodani, Chuo-ku, Sagamihara, Kanagawa 252-5210 Japan}

\keywords{Galaxy: center --- ISM: kinematics and dynamics --- (ISM: evolution) --- (ISM: cosmic-rays)}

\begin{abstract}
This letter presents a Nyquist-sampled, high-resolution [\ion{C}{1}] ${^3}P_1$-${^3}P_0$ map of the $-0.2^\circ<\gl<1.2^\circ$ $\times$ $-0.1^\circ<\gb<0^\circ$ region in the Central Molecular Zone (CMZ) taken with the Atacama Submillimeter Telescope Experiment (ASTE) 10 m telescope.
We have found that molecular clouds in the CMZ can be classified into two groups according to their [\ion{C}{1}]/$^{13}\mathrm{CO}$ intensity ratios: a bulk component consisting with clouds with a low, uniform [\ion{C}{1}]/$^{13}\mathrm{CO}$ ratio (0.45) and another component consisting of clouds with high [\ion{C}{1}]/$^{13}\mathrm{CO}$ ratios ($>0.8$).  
The [\ion{C}{1}]-enhanced regions appear in M$-0.02$$-0.07$, the circumnuclear disk, the 180-pc ring and the high velocity compact cloud CO$+0.02$$-0.02$.
We have carried out a large velocity gradient (LVG) analysis and have derived the $\mathrm{C}^0$/CO column density ratio for M$-0.02$$-0.07$ as 0.47, which is approximately twice that of the bulk component of the CMZ (0.26).
We propose several hypotheses on the origin of high $\mathrm{C}^0$ abundance in M$-0.02$$-0.07$, including cosmic-ray/X-ray dissociation and mechanical dissociation of CO in the pre-existing molecular clouds.
We also suggest the possibility that M$-0.02$$-0.07$ is a cloud at an early stage of chemical evolution from diffuse gas, which was possibly formed by the bar-induced mass inflow in the Galactic Center region. 

\end{abstract}

\section{INTRODUCTION}
The Central Molecular Zone (CMZ) in the inner $\simeq$ 200 pc around the Galactic Center is the Milky Way's most active site of massive star formation; the CMZ contains three well-known supermassive clusters and a burst-like star formation region, Sgr B2 \citep{mor96}.
Recent observations have revealed energetic molecular bubbles and high velocity compact clouds (HVCCs) which are also considered to be probes of massive stellar clusters \citep{oka07,tan07,tan09}.  
Many theoretical and observational studies have been conducted to understand how gas is supplied for these cluster formation activities, and hence for the formation of giant molecular clouds (GMCs) wherein massive stars are formed.
\cite{bin91} theorized a model of gas kinematics in the bar-potential in the inner Galaxy, in which the gas inflows from the innermost cusped $x_1$ orbit to the $x_2$ orbits.
The bar-induced gas flow can trigger large scale mass condensation in the $x_1$-$x_2$ orbit-crowding regions and subsequent burst-like cluster formation in Sgr B2 \citep{has94}.
The gas in the $x_2$ orbits is assumed to be transported to the central $\sim$10 pc region possibly by the inner bar of the Galaxy or by other processes \citep{mor96,nam09}.
This secondary gas flow could facilitate the formation of GMCs in the Sgr A complex, the circumnuclear disk (CND) and the massive stellar clusters in the Sgr A and Radio Arc regions \citep{nam09,oka11}.

Atomic carbon ($\Cn$) can be used as an indicator of GMC formation process in the CMZ.
The origin of abundant interstellar $\Cn$ has been a controversial issue and several explanations have been proposed. In terms of chemical evolution, $\Cn$ is thought to be abundant in the early stage of molecular cloud formation \citep{suz92,lee96,mae99}; %
$\Cn$ is mainly present in the interface layer between the atomic and molecular phases in the photodissociation regions \citep[PDRs;][]{HT99}, but it is also abundant in the inner regions of young molecular clouds because $\Cn\rightarrow$CO conversion requires a timescale of the order of Myr \citep{suz92,lee96}.
This timescale is comparable to the dynamical timescales of  molecular clouds, and hence one can expect molecular cloud formation regions to be clearly visible owing to their high $\Cn$ abundances.

\cite{jaf96}, \cite{ohj01} and \cite{mar04} conducted surveys of the submillimeter [\ion{C}{1}] emission toward the CMZ. 
\cite{ohj01} found that the [\ion{C}{1}] $\CIa$/$\COt\,\JJ{1}{0}$ ratio was uniform throughout the CMZ, whereas \cite{jaf96} showed an increase in the ratio for the inner 6 pc region of the CMZ.
However, these studies were conducted with insufficient spatial resolution or limited spatial coverage, and hence the spatial variation in the $\Cn$ abundance in the CMZ was not investigated in detail.
In this paper, we present a new high-resolution [\ion{C}{1}] $\CIa$ map of the CMZ, and report  the discovery of molecular clouds with high $\Cn$ abundance.

\section{OBSERVATIONS} 

We carried out mapping observations of the CMZ in the [\ion{C}{1}] $\CIa$ (492.1607 GHz) line by using the Atacama Submillimeter Telescope Experiment \citep[ASTE;][]{ezw04} 10 m telescope in October and November 2010.
As a front-end we used the ALMA band 8 QM receiver \citep{sat08}. 
The telescope beam size was $17''$ at 500 GHz.
The digital backend was operated in the wide-band mode with a channel width of 512 kHz.
The typical system noise temperature during the observations was 2000--3000 K. 

We performed on-the-fly (OTF) scans covering the $-0.2^\circ < \gl <  1.25^\circ \times -0.1^\circ < \gb < 0^\circ$ region. 
The reference position was taken at $(\gl, \gb)$=$(1^\circ, -1^\circ)$.
The antenna pointing was checked by CO $\JJ{4}{3}$ observations toward V1427 Aql, and the pointing accuracy was maintained within $5 ''$.
The data was formed into an $\gl$-$\gb$-$\vlsr$ data cube with a $17''\times17''\times 2\ \kmps$\ grid and a $34''$ angular resolution.
The antenna temperatures were calibrated using a standard chopper-wheel method, and were then corrected for the main-beam efficiency of $0.50$.
The estimated rms intensity calibration error was 8 \%. 
The total on-source integration time was 8 hours, thus giving an rms noise level of 0.3 K in the $\Tmb$ scale.
The intensity scale was checked by comparing our data with the data obtained using the CSO telescope \citep{ser94}.
The peak intensity at $(\gl, \gb)=(-0.056^\circ, -0.045^\circ)$ was $7.0\pm0.3$ K in our data, which was in good agreement with the intensity of 7 K in the CSO data.

\section{RESULTS}
\subsection{[\ion{C}{1}]-enhanced Regions}
\begin{figure*}[p]
\begin{center}
\epsscale{.9} 
\plotone{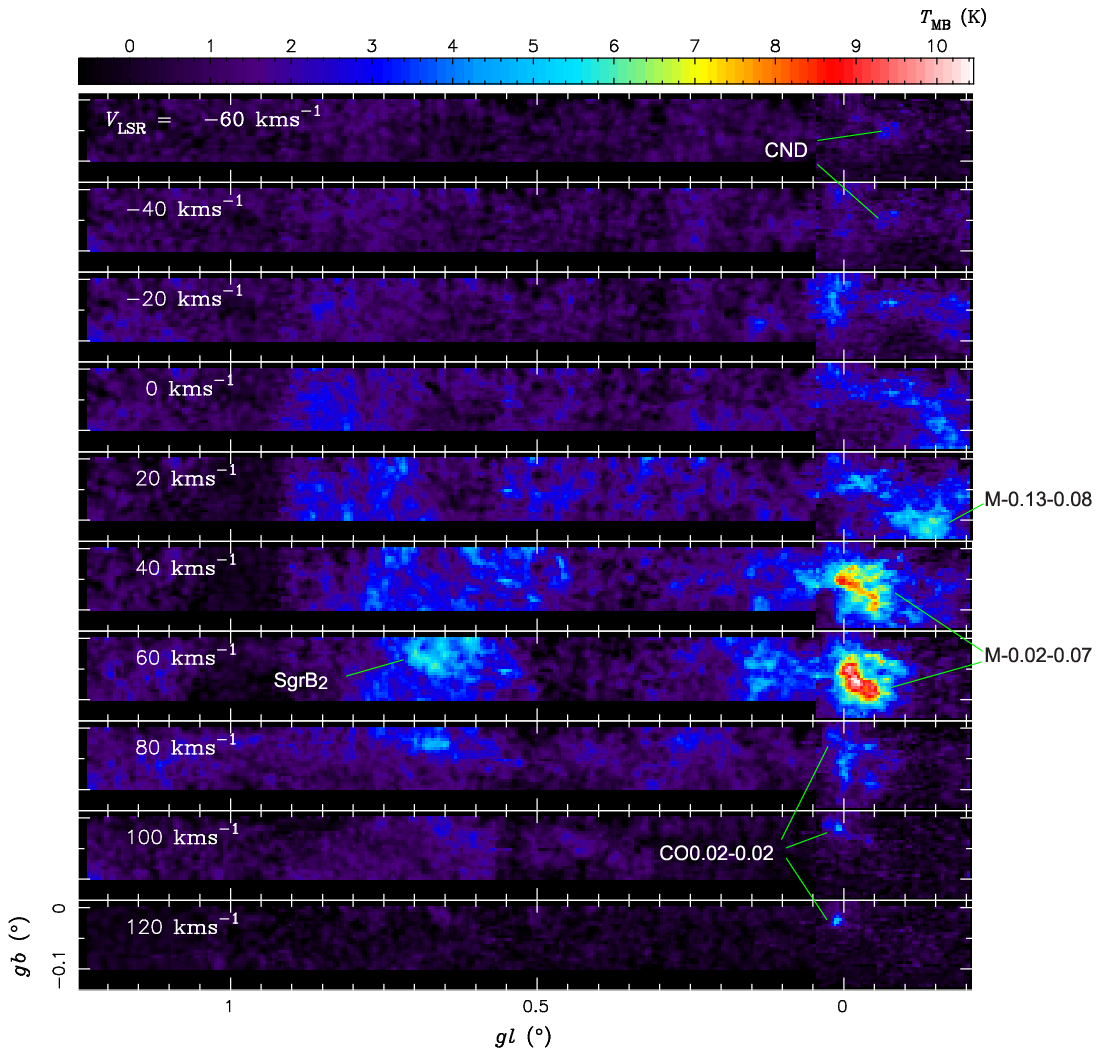}
\caption{Velocity channel maps of [\ion{C}{1}] $\CIa$.
}\label{FIG1}
\end{center}
\end{figure*}

Figure \ref{FIG1} shows the velocity-channel maps of the [\CI] line in a velocity range from $-60$ to $120\ \kmps$.
By comparing the maps with the $\COt\ \JJ{1}{0}$ map \citep{oka98}, we observe several regions with high [\CI]/$\COt$ intensity ratio.
M$-0.02$$-0.07$ (the 50 $\kmps$ cloud) has a [\CI] peak intensity that is approximately twice those of M$-0.13$$-0.08$ (the 20 $\kmps$ cloud) and Sgr B2, whereas these three GMCs have similar peak $\COt$ intensities \citep{oka98}.
Another notable observation is the strong [\CI] emission from CO$+0.02$$-0.02$ in high velocity channels ($\vlsr \ge 100\ \kmps$).
CO$+0.02$$-0.02$ is one of the most energetic HVCCs in the CMZ \citep{oka08}.
Despite its weak detection in the $\COt$ map, the peak [\CI] intensity of this cloud is very high ($5.5$ K) and is comparable to the [\CI] intensity in Sgr B2.

\begin{figure*}[p]
\epsscale{1.}
\plotone{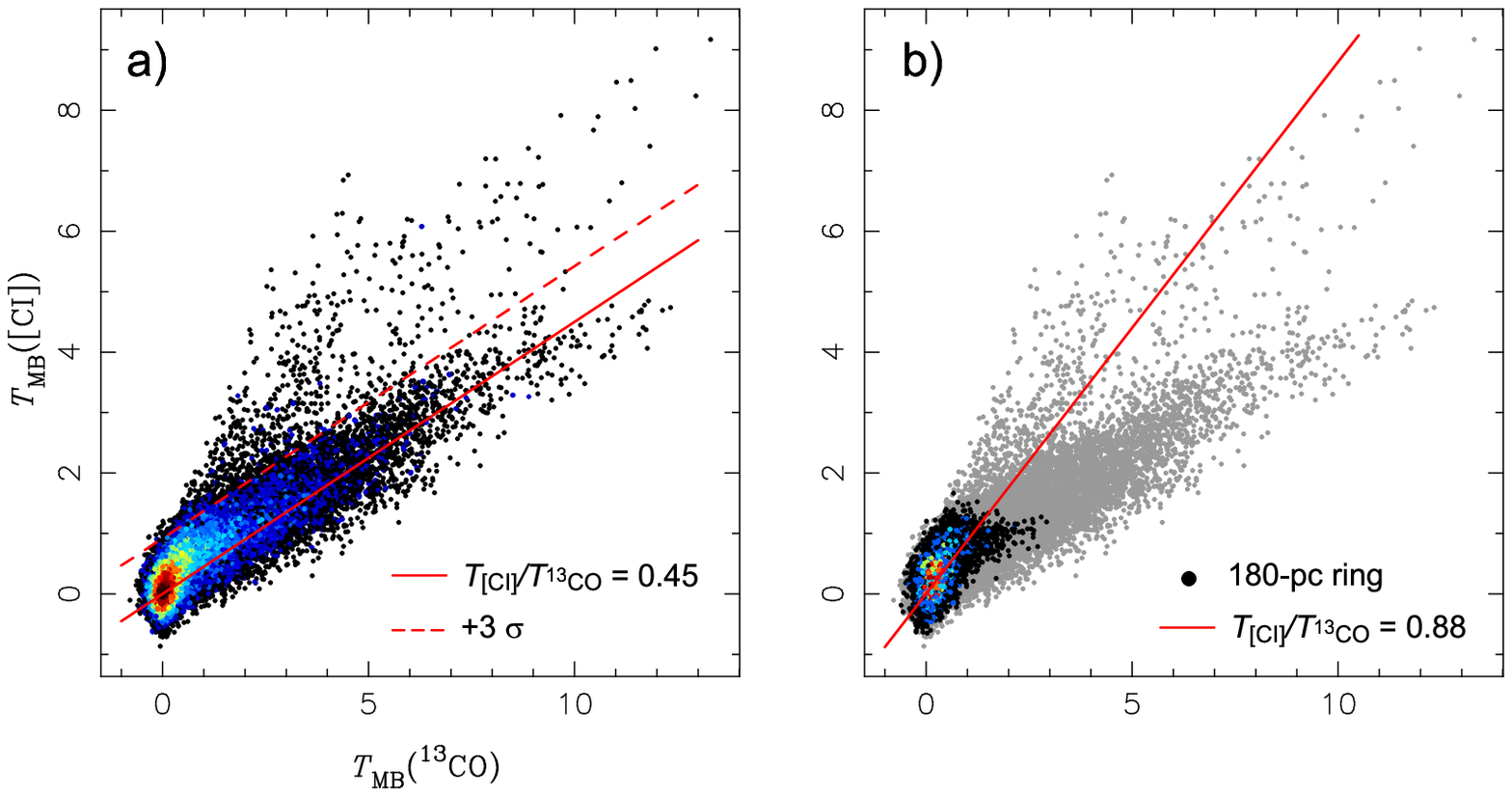}
\caption{(a) $\TCI$-$\TCOt$ scatter plot.  The [\CI] and $\COt$ data are smoothed to $60''\times60''\times4\,\kmps$ resolution. Only independent pixels are plotted.  (b) same as (a), but the 180-pc ring is plotted with colored points.}\label{FIG2}
\end{figure*}

We made a scatter plot of the $\COt$ intensity versus the [\CI] intensity ($\TCOt$ and $\TCI$, respectively).
The [\CI] data were convolved with a $60''$ beam and regridded to match the resolution of the $\COt$ data. 
The velocity resolution of the [\CI] and $\COt$ data was also reduced to $4\ \kmps$ in order to improve the signal-to-noise ratio.
The scatter plot shown in Fig.\ref{FIG2} clearly indicates the presence of [\CI]-enhanced region in the CMZ.
The data points are divided into two components according to their $\TCI$/$\TCOt$ ratios: a bulk component with a uniform $\TCI$/$\TCOt$ ratio of 0.45, and another component with a ratio approximately twice that of the bulk component.

\begin{figure*}[p]
\epsscale{0.75}
\plotone{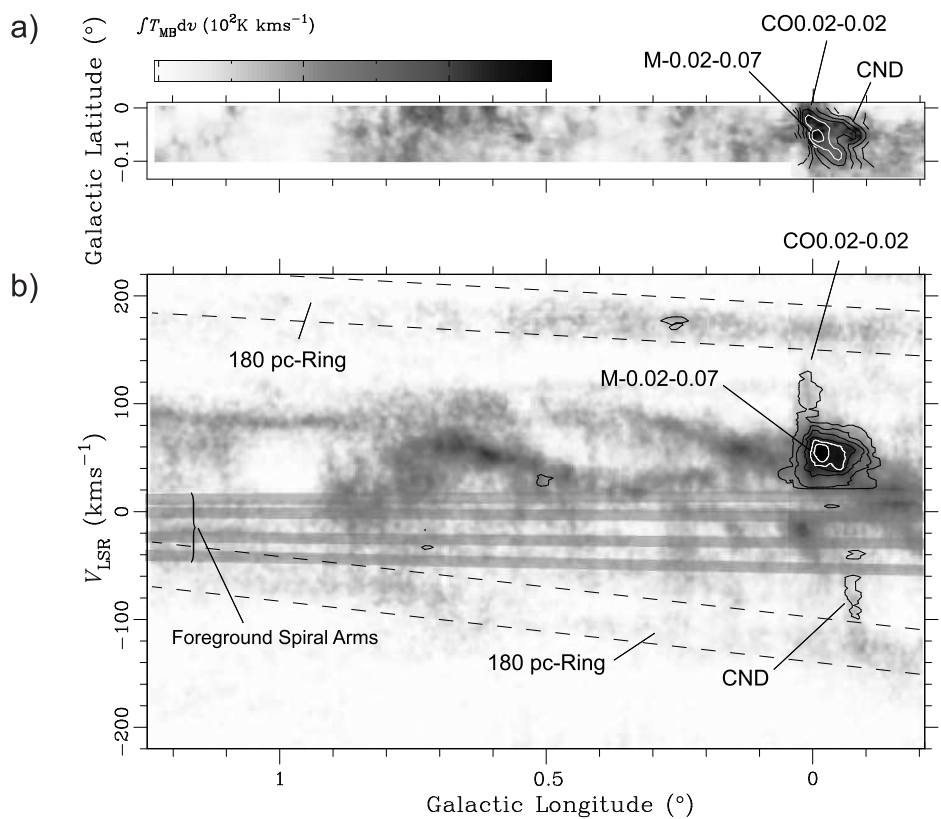}
\caption{(a) Velocity-integrated [\CI] intensity of [\CI]-enhanced region (contours) overlying total velocity-integrated intensity of [\CI] (grayscale).  The contours are drawn at 50 K $\kmps$ intervals.   (b) same as (a), but in longitude-velocity diagram.  The contour levels are 0.1, 1, 2, 3, 4 and 5 K. 
}\label{FIG3}
\end{figure*}

We extracted clouds that belong to the latter, [\CI]-enhanced component according to the following criteria: (1) $\TCI > 0.45\times\TCOt + 3 \sigma$, where $\sigma$ is the noise level of $\TCI$, and (2) the size of the high $\TCI$ region is greater than one resolution element in each of the $\gl$, $\gb$, and $\vlsr$ directions. 
Figure \ref{FIG3} shows the distribution of the [\CI]-enhanced regions in the velocity-integrated intensity map and the $\gl$-$\vlsr$ diagram.
The [\CI]-enhanced regions correspond to three clouds in Sgr A: M$-0.02$$-0.07$, CO$+0.02$$-0.02$ and the CND. 
An increase in $\TCI$/$\TCOt$ ratio for the Sgr A complex has been observed in the low-resolution data of \cite{ohj01}, and our results show that the increase is mainly due to the contribution from M$-0.02$$-0.07$.   
The Sgr A complex has another massive GMC M$-0.13$$-0.08$, but this cloud has a typical, low $\TCI$/$\TCOt$ ratio.

Figure \ref{FIG2}b shows that the 180-pc ring also has a high $\TCI/\TCOt$ ratio, although most part of the ring was not identified as a [\CI]-enhanced region according to the above criteria owing to its low [\CI] intensity.
The best-fit value for the $\TCI$/$\TCOt$ ratio was 0.88. 

\subsection{C$^0$/CO Abundance Ratio}\label{LVG}
We carried out a large velocity gradient (LVG) analysis to estimate the $\NCn/\NCO$ column density ratio on the basis of the $\TCI/\TCOt$ ratio.
For the bulk component, the $\NCn/\NCO$ ratio was calculated as 0.26, assuming both the [\CI] and $\COt$ lines to be optically thin, $\Tkin=50\ \kelvin$ and $\nH=10^{3.5}\ \pcc$ \citep{mar04,nag07}. 
The C/$\Ct$ isotopic ratio was assumed to be 24 according to \cite{lp90}.  
In a typical CMZ environment, $\TCOt$ decreases with increasing excitation temperature of the $\COt$ line, whereas $\TCI$ is insensitive to the physical conditions. 
Hence, the enhanced $\TCI$/$\TCOt$ ratios of M$-0.02-0.07$, CO$+0.02-0.02$, the CND and the 180-pc ring can be attributed either to high $\Cn$ abundance or to high excitation temperature of $\COt$. 

The $\TCI/\TCOt$ ratio averaged over M$-0.02$$-0.07$ was 0.87, which is likely to be caused by the enhanced $\NCn/\NCO$ ratio.
Otherwise, the $\TCI/\TCOt$ ratio of 0.87 would require $\Tkin=370\ \kelvin$ when $\nH=10^{3.5}\ \pcc$ or $\nH=10^{4.6}\ \pcc$ when $\Tkin=50\ \,\kelvin$; however, it is unlikely that the entire cloud has temperature or density of an order higher than the CMZ average.   
On the basis of a multi-transition study on CO and $\COt$, \cite{mar04} estimated that $\Tkin=50$--$60\ \kelvin$ and $\nH=10^{3.6}\ \pcc$ at the center of M$-0.02$$-0.07$.
\cite{min05} showed that the $\COt\ \JJ{2}{1}$ excitation temperature was 15--30 K for both M$-0.02$$-0.07$ and M$-0.13$$-0.08$, and these values are in good agreement with those of \cite{mar04}.
We calculated the $\NCn$/$\NCO$ ratio as 0.47 by assuming that $\Tkin=60\ \kelvin$ and $\nH=10^{3.6}\ \pcc$.

The $\TCI$/$\TCOt$ ratios for CO$+0.02$$-0.02$ and the CND were 2.0 and 2.7, respectively; however, it is not clear whether they have enhanced $\Cn$ abundances because we could not estimate their $\COt$ excitation temperatures accurately.
The $\Tkin$ and $\nH$ values were measured to be $\Tkin=63 \mbox{--} 450\ \kelvin$ and $\nH=10^{4.1\mbox{--}7} \pcc$ \citep[ and references therein]{oka11}.
The $\TCI$/$\TCOt$ ratio of 2.7 for the CND can be explained by the reasonable assumption that $\Tkin=160\,\kelvin$ and $\nH=10^{6.5}\ \pcc$ even if the $\NCn$/$\NCO$ ratio is assumed to be the same as that for the bulk component.
For CO$+0.02$$-0.02$, \cite{oka08} estimated that $\Tkin \gtrsim60\ \kelvin$ and $\nH \gtrsim10^{4.2}\ \pcc$, which yields $\NCn/\NCO \lesssim0.68$.

The temperature and density of the 180-pc ring were estimated as $\sim30\ \kelvin$ and $\sim10^{3.0\mbox{--}3.5}\ \pcc$, respectively \citep{nag07}.
The $\NCn$/$\NCO$ ratio was then estimated to be 0.61--0.77 by assuming $\mathrm{C}/\mathrm{C}^{13}=24$. 
However, this $\NCn$/$\NCO$ ratio may be overestimated because a higher C/$^{13}$C isotopic ratio of $\gtrsim40$ has been suggested for the ring \citep{riq11}.
The $\NCn$/$\NCO$ ratio decreases to 0.37--0.47 if we adopt 40 instead of 24 as the C/$^{13}$C ratio. 

\section{DISCUSSIONS}

\subsection{M$-0.02$$-0.07$}

\begin{figure*}[ttt]
\epsscale{0.8}
\plotone{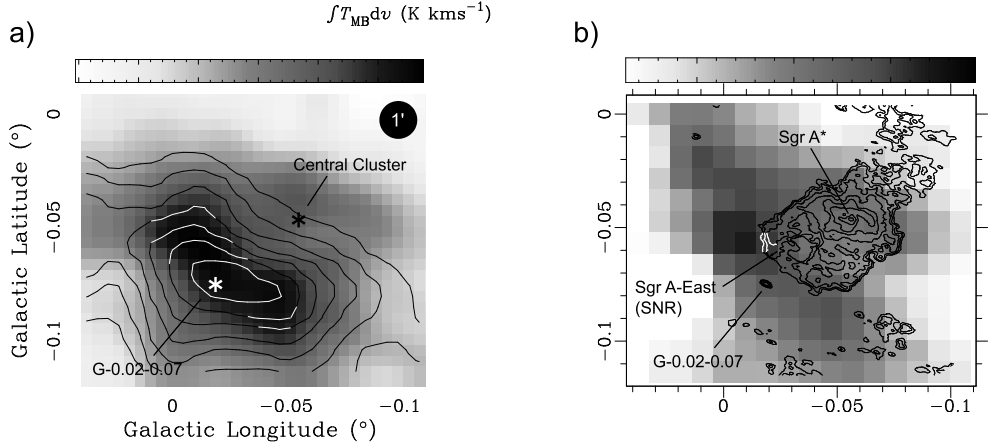}
\caption{(a) Velocity-integrated [\ion{C}{1}] intensity in $\vlsr$ range of 30 to 70 $\kmps$ (grayscale) in Sgr A, smoothed to $60''$ resolution. Contours of $\COt\ \JJ{1}{0}$ are drawn at 50 K  $\kmps$ intervals beginning from 150 K $\kmps$.
(b) [\CI] integrated intensity of [\CI]-enhanced regions, with overlaid contours of 6 cm continuum extracted from Very Large Array (VLA) archival data.
}\label{FIG4}
\end{figure*}

\begin{figure*}[ttt]
\epsscale{0.7}
\plotone{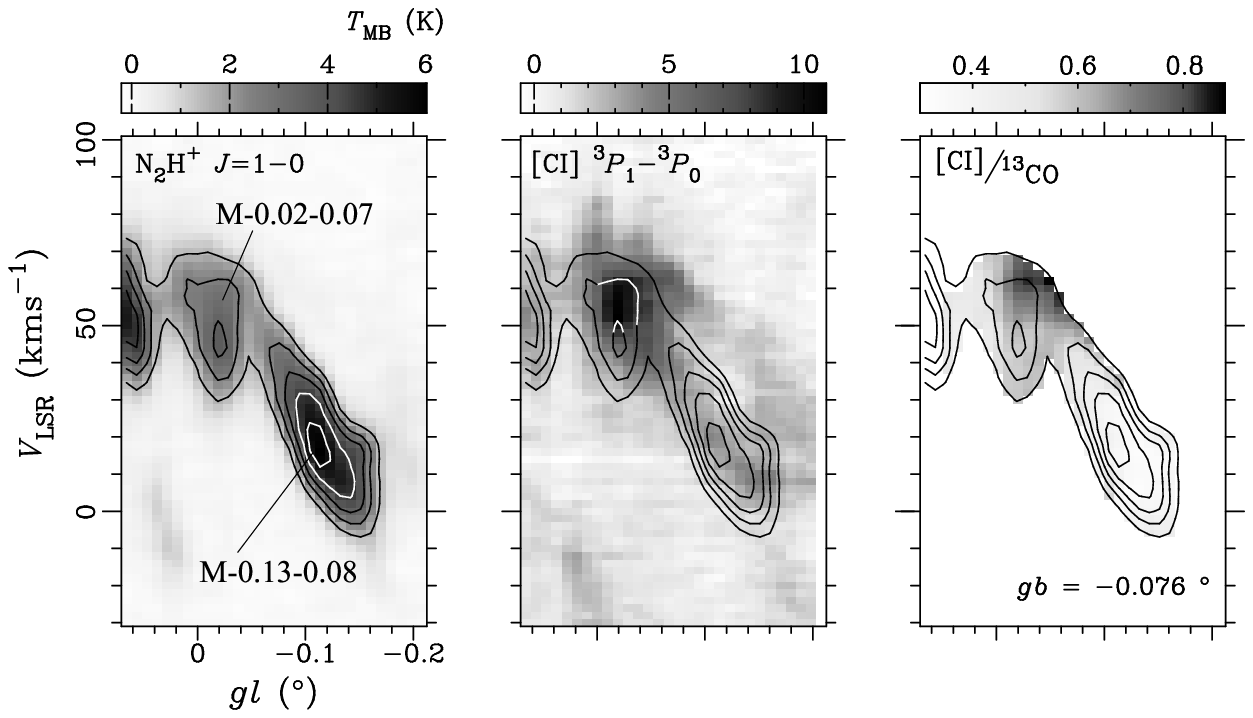}
\caption{Longitude-velocity diagrams of $\NNHp$, [\ion{C}{1}], and $\TCI$/$\TCOt$ ratio in Sgr A at $\gb=-0.076^\circ$.  Contours of the $\NNHp$ intensity are drawn at 1 $\kelvin$ intervals beginning from 1.5 $\kelvin$. }  \label{FIG5}
\end{figure*}

M$-0.02$$-0.07$ is the largest [\CI]-enhanced region in our data sets.
The $\NCn/\NCO$ ratio for the cloud is 0.47, which is approximately five times the typical value for the Milky Way \citep[0.1:][]{oka05}, and roughly corresponds to the ratio for the central regions in nearby galaxies \citep[0.3--5: ][]{ib02,hit08}.

The $\NCn$/$\NCO$ abundance ratio is a sensitive indicator of molecular cloud formation.
\cite{mae99} found a '[\CI]-rich' molecular cloud in the Taurus region and concluded that the cloud was at an early stage of evolution from the atomic gas.
However, unlike the quiescent dark clouds in the Galactic disk, the GMCs in the CMZ are exposed to dissociative processes besides photodissociation by the interstellar radiation field; they are irradiated by strong ultraviolet (UV) radiation from OB-stars \citep[$G_0 \sim10^3$;][]{rod04} and are possibly exposed to enhanced cosmic-ray/X-ray ionization and dissociative shocks. 
Therefore, in addition to the time-dependent chemical model, we discuss these possible explanations for $\Cn$ overabundance in the following sections. 

\subsubsection{Photodissociation}
In terms of the stationary PDR model \citep{htt91} high $\NCn/\NCO$ ratio can be attributed to an intense UV field; $\NCO$ decreases with increasing UV field strength ($G_0$) whereas $\NCn$ is insensitive to $G_0$.
However, $G_0$ for M$-0.02$$-0.07$ is not remarkably higher than that for other GMCs adjacent to \ion{H}{2} regions. 
The [\ion{C}{2}] $\CIIa$ intensity of M$-0.02$$-0.07$, which is a good measure of $G_0$, is $8\times10^{-4}\ \erg\,\second^{-1}\,\psc\,\str^{-1}$ \citep{pog91}.
This value is within the range of typical values for regions outside Sgr A, $(0.6\mbox{--}1)\times10^{-3}\ \erg\,\second^{-1}\,\psc\,\str^{-1}$ \citep{miz94}.  
A similar [\ion{C}{2}] intensity is also observed for the Radio Arc and Sgr B1 regions \citep{miz94}, where no [\CI] enhancement is observed in our data sets.

In addition, the spatial distribution of the [\CI] emission deviates from the stationary PDR model.   
The [\CI] peak of a photodissociative origin is expected to appear at the cloud surface \citep{htt91,kam03} rather than at the cloud center.
However, this does not correspond to the observed distribution of the [\CI] emission shown in Fig.\ref{FIG4}.
The strong [\CI] emission originates from a ridge from $(\gl,\,\gb)=(0^\circ, -0.03^\circ)$ to $(-0.05^\circ, -0.1^\circ)$, coinciding with the $\COt$ ridge.
The [\CI] ridge does not show a significant positional offset from the $\COt$ ridge toward either of the two dominant UV sources, the Central cluster and the G$-0.02$$-0.07$ \ion{H}{2} region.

\subsubsection{Cosmic-ray/X-ray Ionization}
Chemical models show that a high ionization rate due to cosmic-rays or X-rays increases $\Cn$ abundance \citep{flo94,mei06}.
In fact, very high $\NCn$/$\NCO$ ratios ($\gtrsim1$) are found for the central regions of starburst galaxies and active nuclei \citep{ib02,hit08} and for GMCs interacting with supernova remnants (SNRs) \citep{whi94,ari99}, where high cosmic-ray/X-ray flux is expected.

M$-0.02$$-0.07$ is located near Sgr A$^*$, which is considered as a possible source of cosmic-rays in the CMZ \citep{che11}.
It is also argued that Sgr A$^*$ underwent a strong X-ray outburst in the recent past \citep{koy96}.
In addition, M$-0.02$$-0.07$ interacts with the Sgr A-East SNR \citep{yz01}, which is another possible cosmic-ray source in the Sgr A region. 
The cosmic-ray/X-ray dissociation by these sources may explain the high $\Cn$ abundance in M$-0.02$$-0.07$.
An advantage of the cosmic-ray/X-ray dissociation model over the standard PDR model is that it can explain the spatial co-existence of the $\COt$ and [CI] emissions more easily, because the X-rays and cosmic-rays can penetrate deeper into a cloud than UV photons. 

\subsubsection{Mechanical Dissociation}
Propagation of a fast shock thorough dense molecular gas can dissociate CO in the post shock gas \citep{hm80}.
\cite{whi94} suggested that the very high $\Cn$/CO ratio ($1.3$--$2.9$) for the IC443C C-shocked region can be attributed to the blast wave from the SNR, or to the enhanced cosmic-ray flux in the SN-shocked region.  

The dissociative shock from Sgr A-East may provide an alternative explanation for the high $\Cn$/CO ratio, especially at the Galactic northern and western edge of the SNR shell where the shape of the [\CI]-enhanced region spatially correlates well with that of the SNR (Fig.\ref{FIG4}b). 
Another possible source of the dissociative shock is cloud-cloud collision.
It is suggested that  cloud-cloud collision is rather frequent in the CMZ because of the high volume filling factor of molecular clouds and the presence of a bar-potential \citep{has94,hue98}.

\subsubsection{Time Dependent Chemistry}
The high $\Cn$ abundance of M$-0.02$$-0.07$ can be also explained by time-dependent chemical models.
The models of \cite{suz92}, \cite{lee96}, and \cite{ber97} showed that molecular clouds are $\Cn$-abundant for $\sim1$ Myr after their formation.
Hence, M$-0.02$$-0.07$ region can be understood as a young molecular cloud similar to the [\CI]-rich cloud in the Taurus region \citep{mae99}.

In fact, the [\CI]-enhanced region in M$-0.02$$-0.07$ appears spatially separated from the evolved, star-forming dense core region of the cloud.
In Fig.\ref{FIG5}, the distribution of the [\CI] emission is compared to that of the $\NNHp \JJ{1}{0}$ line (Oka et al. private communication), which is a tracer of evolved dense cores.
M$-0.02$$-0.07$ is remarkably weak in $\NNHp$ as compared to the neighboring GMC, M$-0.13$$-0.08$; the $\NNHp\,\JJ{1}{0}$/$\HCNt\, \JJ{1}{0}$ ratio for M$-0.02$$-0.07$ is 0.59, which is two times lower than that in M$-0.13$$-0.08$, 1.1 (Oka et al. private communication).
Further, we note a negative spatial correlation between [\CI] and $\NNHp$ in the internal structure of M$-0.02$$-0.07$.
The [\CI] peak velocity of M$-0.02$$-0.07$ is 50--65 $\kmps$, whereas the $\NNHp$ peak velocity is 45--50 $\kmps$.
The $\TCI/\TCOt$ ratio at the $\NNHp$ peak is 0.54, which is significantly lower than that averaged over the entire cloud.
Since $\NNHp$ becomes abundant in the later phase ($\gtrsim1$ Myr) of chemical evolution \citep{hir95,ber97}, the observed negative spatial correlation between $\NNHp$ and [\CI] is consistent with the chemical evolution scenario.

\subsection{Mass Inflow in the CMZ}
If the [\CI]-enhanced region in M$-0.02$$-0.07$ is a young molecular cloud with an age not much greater than the chemical timescale of $\Cn\rightarrow$CO conversion, supply of a large amount of diffuse gas in the past $\sim1\ \mathrm{Myr}$ would be required for its formation.
From the mass of M$-0.02-0.07$ \citep[$\sim10^5\ \Msol$; ][]{zyl90}, the supply rate is estimated as $\sim0.1\ \Msol\,\mathrm{yr}^{-1}$, which is within a reasonable range that could be explained by the mass inflow rates specified in the literature. 
\cite{mor96} estimated the inflow rate to be 0.1--1 $\Msol\ \mathrm{yr}^{-1}$ at $\sim200$ pc from the Galactic Center.
\cite{nam09} argued that M$-0.02$$-0.07$ is a part of a gas disk formed by the mass inflow to the central $\sim15$ pc region driven by the inner bar potential.
The mass inflow rate estimated by their model is $\sim0.1\ \Msol$ yr$^{-1}$, which is also in good agreement with our estimate.

The high $\TCI/\TCOt$ ratio for the 180-pc ring can be also explained in the framework of the bar-driven inflow model.
\cite{bin91} showed that the 180-pc molecular ring is formed by shock compression of the atomic gas at the inner edge of the innermost $x_1$ orbit and that the molecular gas rapidly flows into the inner $x_2$ orbits.
The residence time of the molecular gas in the ring is shorter than the orbital period of the ring,  $\simeq\pi\,R_{\mathrm{ring}} / V_{\mathrm{ring}} \sim6$\,Myr, where $R_{\mathrm{ring}}$ and $V_{\mathrm{ring}}$ are the radius and the rotational velocity of the ring, respectively.
This timescale is comparable to, or slightly longer than the chemical timescale of $\Cn\rightarrow\CO$ conversion, and hence a considerable fraction of the molecular gas in the ring may be $\Cn$-abundant.

However, we note that it is difficult to draw a definite conclusion about the origin of the [\CI]-enhanced regions because of the complexity of the Galactic Center environment.
As discussed in the previous subsections, cosmic-ray/X-ray dissociation or the mechanical dissociation of CO in the pre-existing molecular clouds can also provide reasonable explanations for the high $\NCn/\NCO$ ratio.
For the 180-pc ring, we should also consider the possibility that the high [\CI]/$\COt$ intensity ratio may not be due to the high $\Cn$ abundance but to the low $\Ct$ isotopic abundance.

\subsection{CND and CO$+0.02$$-0.02$}
The CND has the highest $\TCI/\TCOt$ ratio of the clouds in our data sets, although we could not confirm the increase in its $\Cn$ abundance. 
\cite{oka11} found the CND has low $\NNHp$ abundance, similar to M$-0.02$$-0.07$.
This result, along with the high $\TCI/\TCOt$ ratio, suggests similarities in the chemical composition of these clouds.

The $\TCI/\TCOt$ ratio of 2.0 for CO$+0.02$$-0.02$ is considerably higher than that for M$-0.02$$-0.07$.
The large velocity width of CO$+0.02$$-0.02$ indicates that the cloud is violently shocked, although the driving source of the shock is not identified.
\cite{oka08} argued that the energetic internal motion of CO$+0.02$$-0.02$ is driven by a series of SN explosions.
The effect of the shock dissociation, and possibly of cosmic-ray dissociation enhanced by the SN-shock, may be more important for C$+0.02$$-0.02$ than for M$-0.02$$-0.07$.

\section{SUMMARY}
We report the discovery of molecular clouds with high $\Cn$ abundance in the CMZ.
We found that the $\TCI/\TCOt$ ratio significantly increased for M$-0.02$$-0.07$, CO$+0.02$$-0.02$, the CND, and the 180-pc ring, as compared to that for the bulk component of the CMZ.
The $\NCn$/$\NCO$ ratio of 0.47 for M$-0.02$$-0.07$ is approximately twice the CMZ average.

We could not draw a definite conclusion on the origin of the high $\NCn/\NCO$ ratio because of the complexity of the Galactic Center environment.
We propose  cosmic-ray/X-ray ionization and mechanical dissociation by fast shock as possible explanations.
We also hypothesize that the [\CI]-enhanced regions in M$-0.02$$-0.07$ and the 180-pc ring are young molecular clouds with ages not greater than the chemical timescale of $\Cn\rightarrow$CO conversion.
Such young, massive molecular clouds were possibly formed by the bar-induced mass inflow in the Galactic Center region.


\end{document}